# Albert Einstein: Rebellious Wunderkind

Galina Weinstein

Childhood and Schooldays: Albert Einstein, and the family members seemed to have exaggerated the story of Albert who developed slowly, learned to talk late, and whose parents thought he was abnormal. These and other stories were adopted by biographers as if they really happened in the form that Albert and his sister told them. Hence biographers were inspired by them to create a mythical public image of Albert Einstein. Albert had tendency toward temper tantrums, the young impudent rebel Einstein had an impulsive and upright nature. He rebelled against authority and refused to learn by rote. He could not easily bring himself to study what did not interest him at school, especially humanistic subjects. And so his sister told the story that his Greek professor, to whom he once submitted an especially poor paper, went so far in his anger to declare that nothing would ever become of him. Albert learned subjects in advance when it came to sciences; and during the vacation of a few months from school, Albert independently worked his way through the entire prospective Gymnasium syllabus. He also taught himself natural science, geometry and philosophy by reading books that he obtained from a poor Jewish medical student of Polish nationality, Max Talmud, and from his uncle Jacob Einstein.

## 1. Albert was not a good orator

In 1924 in her *Biographical Sketch* (after Einstein became world famous), Einstein's sister, Maja, told the following story:[1]

Albert as a child "would play by himself for hours. […] he developed slowly in childhood, and he had such difficulty with language that those around him feared he would never learn to speak. But this fear also proved unfounded".

In 1930 Anton Reiser, also recounted the same story, "Slowly, and only after much difficulty, he learned to talk. His parents thought he was abnormal. They hired governess called the still, backward, slow-speaking Albert, 'Pater Langweil' (Father Bore)".[2]

The older Einstein also recounted in a letter from 1954: "My parents were worried because I started to talk comparatively late, and they consulted the doctor because of it. I cannot tell how old I was at that time, but certainly not younger than three". Einstein also added: "However, my later development was completely normal except for the peculiarity that I used to repeat my own words softly".[3]

---

[1] Winteler-Einstein Maja, *Albert Einstein –Beitrag für sein Lebensbild*, 1924, reprinted in abridged form in *The Collected Papers of Albert Einstein Vol. 1: The Early Years, 1879–1902* (*CPAE*, Vol. 1), Stachel, John, Cassidy, David C., and Schulmann, Robert (eds.), Princeton: Princeton University Press, 1987, pp xlviii-lxvi; p. Xviii, p. 1vii.
[2] Reiser, Anton, *Albert Einstein: A Biographical Portrait*, 1930/1952, New York : Dover, 1930, p. 27. Rodulf Kayser, the husband of Einstein's stepdaughter Ilse (writing under the pseudonym Anton Reiser).

Maja also reports on this strange linguistic habit. "His early thoroughness in thinking was also reflected in a characteristic, if strange habit. Every sentence he uttered, no matter how routine, he repeated to himself softly, moving his lips. This odd habit persisted until he was seven.[4]

Maja, Albert Einstein, and the family members seemed to have exaggerated the story of Albert who developed slowly, learned to talk late, and whose parents thought he was abnormal. There is no doubt that there is grain of truth to these stories and Maja and Einstein tell in all sincerity their recollections. But these stories sound like family stories, and as such are exaggerated. These and other stories were adopted by biographers as if they really happened in the form that Maja and Albert told them. And thus Biographers were inspired by them to create a mythical public image of Albert Einstein. This widespread image of Einstein embodies stories about Einstein the retarded Genius.

In the above 1954 letter Einstein did go on to say, "Also, I never exactly became an orator later."[5] It appears that indeed it took Einstein quite a long time to become a good lecturer.

**2. Rebel and Creative**

At the age of four or five, young Albert experienced a wonder. His father Herman showed him a compass. This experience, so recounts Einstein himself in his *Autobiographical notes*, changed his life:[6]

"A wonder of this kind I experienced as a child of four or five years when my father showed me a compass. That this needle behaved in such a determined way did not at all fit into the kind of occurrences that could find a place in the unconscious world of concepts (efficacy produced by direct 'touch'). I can still remember – or at least believe I can remember – that this experience made a deep and lasting impression upon me".

And then, "At the age of 12 I experienced a second wonder of a totally different nature: in a little book dealing with Euclidean plane geometry, which came into my hands at the beginning of a schoolyear". The assertions of geometry" and "lucidity and certainty made an indescribable impression upon me".

---

On his 74th birthday Einstein was asked by reporters:[7]

"It is said that you were decisively influenced at the age of five by a compass, and at twelve by a book of Euclidean geometry. Did these things really have any influence on your life work?" Einstein replied: "I myself think so and I believe that these outside influences had a considerable influence on my development. But man has little insight into what goes on within him. When a young puppy sees a compass for the first time it may have no similar influence, nor on many a child. What does, in fact, determine the particular reaction of the individual? One can postulate more or less plausible theories on this subject, but one never really finds the answer".

Einstein had also told the "wonder" compass story to Alexander Moszkowski in 1916,[8]

"His father once showed the infant, as he lay in his cot, a compass, simply with the idea of amusing him – and in the five-year-old boy the swinging metal needle awakened for the first time the greatest wonderment about unknown cohesive forces, a wonderment that was an index of the research spirit that was still lying dormant in his consciousness. The remembrance of this physical event has a significant meaning for the Einstein today. [….] This instrument addressed him in oracular language, indicating to him an electromagnetic field that was in later years to serve him as a domain for fruitful research".

At the age of five or six Albert also received Music lessons on the Violin. Maja said that Albert referred to his music teacher Herr Schmied as Du instead of Sie. "As is well known, in Germany one uses the polite form 'Sie' for adults and for people who are not members of one's family, while 'Du' is used only within the family, among children, and between close friends. There was thus something impertinent, but also something naïve and humorous in little Albert's way of addressing his music teacher with 'Du', Herr Schmied...'".[9]

Until Albert was twelve the Violin lessons gave him no pleasure, remained only a duty as burdensome as school. His musical experience grew out of listening, and pleasure in his playing came but slowly".[10] In 1940, Einstein recounted in a draft of a

---

letter to Philipp Frank: "I took violin lessons from age 6 to 14, but had no luck with my teachers for whom music did not transcend mechanical practicing.[11]

At the age of five-six Albert received his first instruction at home from a woman teacher. Albert's sister Maja, who was probably quite annoyed at her brother who sometimes threw items on her, wrote in 1924 that, Albert inherited from his grandfather Julius Koch a tendency toward violent temper tantrums.[12] Susanne Markwalder from Zurich later spoke of Einstein's "impulsive and upright nature".[13] At such moments", Maja, recalled, "his face would turn completely yellow, the tip of his nose snow-white, and he was no longer in control of himself. On one such occasion he grabbed a chair and struck at his teacher, who was so frightened that she ran away terrified and was never seen again. Another time he threw a large bowling ball at his little sister's head […]. This should suffice", tells us Maja, "to show that it takes a sound skull to be the sister of an intellectual. The violent temper disappeared during his early school years".[14] In fact Einstein's temper tantrums did not disappear.

Brigitte Fischer, the musical daughter of the publisher Samuel Fischer, described Einstein in her autobiography:[15]

"Occasionally I even made music with Albert Einstein, who is known to have been no great musician. He was a great music lover and an impassioned violinist, but he couldn't endure any criticism of his playing and could fly into a rage when something didn't turn out right. One day, he was playing the Bach Double Violin Concerto with an excellent professional violinist, Gerhart Hauptmann's daughter-in-law, and me at the piano. He suddenly broke off and shouted furiously at his partner, whose playing was drowning him out: 'Don't play so loud!' I think he got more excited about that than he ever did in a scientific dispute".

## 3. Einstein cannot wholly withstand authority and obedience

### 3.1 Primary school

In October 1885, so reports Maja, Albert was seven and he entered the public primary school, öffentliche Volksschule.[16] There were no Jewish schools and no secular schools in Munich at that time. Thus Hermann and Pauline registered young Albert to the Volksschule Peterschule: a Catholic elementary school near their home. As the one Jew along with seventy Catholic children he learned catholic lessons and made a good progress in catholic studies. He was even able to assist his classmates.[17]

---

[11] Hoffmann and Dukas, 1973, p. 20; *CPAE*, Vol. 1, editorial note. 39, p. 1viii. Hoffmann and Dukas, 1973, p. 20.
[12] Winteler-Einstein, 1924, *CPAE*, Vol. 1, p. xviii, p. 1vii.
[13] Seelig, 1956, p. 38; Seelig, 1954, p. 44.
[14] Winteler-Einstein, 1924, *CPAE*, Vol. 1, p. xviii, p. 1vii.
[15] Brigitte B. Fischer, *My European Heritage: Life Among Great Men of Letters*, Translated from the German by Harry Zohn, 1986, Boston: Branden Publishing Company, p. 10.
Fischer says that Einstein got more excited about playing violin than he ever did in a scientific dispute. It is not quite true. In 1936 the *Physical Review* rejected one of Einstein's and Nathan Rosen's papers on gravitational waves, provoking a furious reaction: Einstein told the editor he had sent him the paper for publication, and he had not authorized him to show the paper to referees before the paper was printed. After this incident Einstein never published again in *Physical Review*. Kennefick, Daniel, "Einstein versus the *Physical Review*", *Physics Today* 58, 2005, pp. 43-48.
[16] Winteler-Einstein, 1924, *CPAE*, Vol. 1, p. xix, p.1viii.

In 1886, Albert's mother, Pauline, wrote to her mother Jette Koch saying, "Yesterday, Albert got his school marks. Again he is at the top of his class and got a brilliant record".[18] Maja wrote that, Albert had a rather strict teacher whose methods included teaching children arithmetic, and especially the multiplication tables, with the help of whacks on the hands, so-called beats; "a style of teaching", so explains Maja, "that was not unusual at the time, and that prepared the children early for their future role as citizens".[19]

"On the whole", says Philipp Frank, "Einstein felt that school was not very different from his conception of barracks – that is, a place where one was subject to the power of an organization that exercised a mechanical pressure on the individual, leaving no area open within which he might carry on some activity suited to his nature. The students were required to learn mechanically the material presented to them, and the main emphasis was placed on the inculcation of obedience and discipline".[20]

Frank explained that, "It was very characteristic of young Einstein's religious feeling that he saw no noticeable difference between what he learned of the Catholic religion at school and the rather vaguely remembered remnants of Jewish tradition with which he was familiar at home".[21] This is in fact very characteristic of the older tolerant and humanistic Einstein, speaking with Frank while the latter wrote the biography in the 1940's.

Maja said that, when Albert entered public school, his religious instruction, then compulsory in Bavaria, also had to begin. Maja tells that "a liberal spirit, and non-dogmatic in matters of religion, brought by both parents from their respective homes, prevailed within the family. There was no discussion of religious matters or rules at the Einstein home. But since Albert was obliged to receive religious instruction, he was taught at home by a distant relative".[22]

Reiser told the story of the teacher bringing a nail one day to the class,[23]

"The Catholic teacher of religion liked him. But one day the same teacher brought a large nail to class and told the pupils that it was the nail with which the Jews had nailed Jesus to the cross. The incident stimulated in the pupils anti-Semitic feeling which was turned against their Jewish fellow-student Einstein. For the first time Albert experienced the frightful venom of anti-Semitism".

---

[17] Winteler-Einstein, 1924, *CPAE*, Vol. 1, p. xix, p. 1viii.
[18] Hoffmann and Dukas, 1973, p. 19.
[19] Winteler-Einstein, 1924, *CPAE*, Vol 1, p. xix, p. 1viii.
[20] Frank, Philip, *Einstein: His Life and Times*, 1947, New York: Knopf, 2002, London: Jonathan, Cape, p. 10; Frank, Philip, *Albert Einstein sein Leben und seine Zeit*, 1949/1979, Braunschweig: F. Vieweg, p. 23.
[21] Frank, 1947/2002, p. 10; Frank, 1949/1979, p. 23.
[22] Winteler-Einstein, 1924, *CPAE*, Vol. 1, p. xx, p.1ix.
[23] Reiser, 1930, p. 30.

Frank wrote that the teacher told the pupils, "The nails with which Christ was nailed to the cross looked like this". But the teacher did not add, as sometimes happens, that the Crucifixion was the work of the Jews. Nor did the idea enter the minds of the students that because of this they must change their relations with their classmate Albert. Nevertheless, according to Frank, Einstein found this kind of teaching rather uncongenial, because it recalled the brutal act connected with it, and because he sensed it awakens latent sadistic tendencies.[24]

Max Jammer explains in his book, *Einstein and Religion*,[25]

"Frank's biography is known to be based largely on epistolary correspondence, whereas Kayser's account is based on personal conversations with Einstein. In his brief preface to Kayser's biography, Einstein declared, 'I found the facts of the book duly accurate […]'. It is, of course difficult today to find out which of the two versions is true. It is also difficult to assess how such an anti-Semitic incident, had it really happened, would have affected Albert's religious attitude toward Judaism".

Moszkowski wrote, "At the age of eight or nine he presented the picture of a shy, hesitating, unsociable boy, who passed on his way alone, dreaming to himself, and going to and from school without feeling the need of a comrade. He was nicknamed "Biedermaier" (Honest John), because he was looked on as having a pathological love for truth and justice".[26] The pathological love for truth ("Biedermaier" trait) was also described later in 1952 by Einstein's former classmate Hans Byland from the secondary school, Aargau Kantonsschule.[27]

Byland, a year older than Einstein, had painted Einstein's portrait as a teenage in the following words: "The great physicist as a young man could not be fitted into any pattern".[28] Einstein was "impudent Swabian [...] Sure of himself, his grey felt hat pushed back on his thick, black hair, he strode energetically up and down in the rapid, I might almost say crazy, tempo of a restless spirit which carries a whole world in itself. Nothing escaped the sharp gaze of his large bright brown eyes. Whoever approved him immediately came under the spell of his superior personality. A sarcastic curl of his rather full mouth with the protruding lower lip did not encourage Philistines to fraternize with him". His friend adds that "his attitude towards the world was that of a laughing philosopher and his witty mockery pitilessly lashed any conceit or pose. In conversation he always had something to give. […] He made no bones about voicing his personal opinions whether they are offended or not. This courageous love of truth gave his whole personality a certain cachet which, in the long run, was bound to impress even his opponents".[29]

---

[24] Frank, 1947/2002, p. 10; Frank, 1949/1979, pp. 22-23.
[25] Jammer, Max, *Einstein and Religion Physics and Theology*, 1999, Princeton: Princeton University Press, p. 21.
[26] Moszkowski, 1921a, p. 222; Moszkowski, 1921b, p. 220. "Man gab ihm den Spitznamen 'Biedermaier', weil man ihn für krankhaft wahrheits- und gerechtigkeitsliebend hielt".
[27] Seelig, 1956, p. 14; Seelig, 1954, p. 17.
[28] Seelig, 1956, p. 13; Seelig, 1954, p. 15.

The aging Byland of 1952 probably forgot that when he had studied with the young impudent rebel Einstein (who was not yet famous), and the latter voiced his personal opinions (whether they were offensive or not), his opponents were not impressed at all…When Einstein was nine years old, because of his conscientiousness in not making any false statements or telling lies he was called *Biedermeier* (Honest John) by his classmates. At the age of nine Einstein was Honest John and did not tell lies. He told the truth, and later the rebel Einstein thought the most important thing was to tell the truth, and thus he voiced his personal opinions whether they were offending or not.

Moszkowski reported in his book about Einstein's impression of his school,[30]

"He told me with bitter sarcasm that his teachers had the character of sergeants – those latter in the gymnasium (secondary school) were of the nature of lieutenants. Both terms are used in the pre-armistice sense, and his words were directed against the self-opinionated tone and customs of these garrison-schools of earlier days".

Frank reported exactly the same story relying on Moszkowski.[31] Einstein wrote Frank in 1940 about the spiritless and mechanistic method of teaching at school, which seemed pointless when he confronted difficulties because of his poor memory for words. Einstein wrote Frank that he preferred to accept any kind of punishment as he tried to learn by heart at school.[32]

Einstein's biographer, Carl Seelig reported: "The worst of all, in my view", so Einstein said later, "is when a school is mainly run by fear, power and artificial authority. Such treatment destroys the healthy feelings, the integrity and self-confidence of the pupils. All that it produces is a servile helot".[33]

---

[29] Seelig, 1956, p. 14; Seelig, 1954, pp. 16-17.
[30] Miszkowski, 1921a, p. 223; Miszkowski, 1921b, p. 221.
[31] Frank, 1947/2002, p. 11; Frank, 1949/1979, pp. 24-25.
Frank explained that the sergeants in the German army of Wilhelm II were notorious for their coarse and often brutal behavior toward the common soldiers. On the other hand, the lieutenants were members of the upper class and did not come into direct contact with the men. They extracted their desire for power in an indirect manner. Thus when Einstein compared his teachers to sergeants and lieutenants, he regarded their tasks to be the enforcing of mechanical order upon the pupils.
[32] *CPAE*, Vol. 1, note 56, p. lxiii.
Frank wrote that, "when Albert was nine years old and in the highest grade of the elementary school, he still lacked fluency of speech, and everything he said was expressed only after thorough consideration and reflection". In the English translation of Frank "(Honest John)" is added in parentheses for the English reader. Frank then says, "No evidence of any special talent could be discovered, and his mother remarked occasionally: 'Maybe he will become a great professor some day, but perhaps she meant only that he might develop into some sort of eccentric". Frank, 1947/2002, p. 10; Frank, 1949/1979, p. 23. Reiser, "Of small Albert, however, she [his mother] often prophesied: 'Some day he will be a great professor'".[32] Recall that in 1886 when Albert was eight Pauline wrote to her mother Jette Koch saying, that Albert got his school marks, and he is at the top of his class and got a brilliant record. Hoffmann and Dukas, 1973, p. 19. Therefore, Albert's mother probably recognized that her son was talented, and it does not seem that she thought he would develop into some sort of eccentric. Frank probably based his above report on one of Maja's stories pertaining to Einstein's early childhood.

**3.2 Secondary School**

On October 1, 1888, at the age of 9.5, Albert entered the Luitpold Gymnasium. The building was almost completely destroyed during world war two. Today the school is called after Albert Einstein, "Gymnasium Albert Einstein", even though as we shall soon see Einstein could not withstand this school. Maja wrote that, in accord with the Gymnasium's humanistic orientation, primary emphasis was placed on classical languages, Latin and later Greek, while mathematics and the natural sciences received less emphasis.[34]

Reiser explained, "[…] as a student of languages, Albert was only mediocre. He lacked the phonetic, as well as the mnemonic, gift. He hated the burden of so much memorizing and did not show the slightest talent for learning by rote, which study of classical languages particularly called for".[35]

Albert the rebel refused to learn humanistic subjects, and refused to learn by rote. Especially, Hoffman and Dukas write that "Einstein could not easily bring himself to study what did not interest him.[36] Einstein was not as good in Latin as he was in natural sciences, and he rebelled against authority; and so Maja told the story that was repeated since then in many biographies: "his Greek professor, to whom he once submitted an especially poor paper, went so far in his anger to declare that nothing would ever become of him. And in fact Albert Einstein never did attain a professorship of Greek grammar".[37] The older Einstein told Carl Seelig the same story that, it was his Latin master who prophesied: "You will never amount to anything, Einstein".[38]

In light of the growing fame of Einstein the genius, this anecdote was indeed incredible, and so biographers grabbed the story. For instance, Clark told the story in his biography in the following way, "a family legend that when Hermann Einstein asked his son's headmaster what profession his son should adopt, the answer was simply: 'It doesn't matter; he'll never make a success of anything'".[39]

---

[33] Seelig, 1956, p. 12; Seelig, 1954, p. 14.
[34] Winteler-Einstein, 1924, *CPAE*, Vol. 1, p. xx, p. 1x. In the *Realschulen* the sciences and natural sciences did receive great emphasis. It is curious that Albert's parents sent him to the *Gymnasium*. In 1955 – the year Einstein died – he attested that, "As a pupil [in Gymnasium] I was neither particularly good nor bad. My principal weakness was a poor memory and especially a poor memory for words and texts". Hoffmann and Dukas, 1973, pp. 19-20.
[35] Reiser, 1030, p. 37.
[36] Hoffmann and Dukas, 1973, p. 28.
[37] Winteler-Einstein, 1924, *CPAE*, Vol. 1, p. xx, pp. 1x-1xi.
[38] "Einstein, aus Ihnen wird nie was Rechtes warden!". Seelig, 1956, p. 12; Seelig, 1954, p. 15.
[39] Clark, Ronald, W., *Einstein The Life and Times*, 1971, New York: The World Publishing Company, p. 10.

However, Einstein was very talented in sciences. Reiser told that, "At fourteen the boy was already master of higher mathematics which the secondary school, based on humanistic principles, did not teach".[40]

In the Gymnasium Albert was supposed to begin the study of algebra and geometry at the age of 13. But by that time he already had a predilection for solving complicated problems in applied arithmetic, although, says Maja, the computational errors he made kept him from appearing particularly talented in the eyes of his teachers. He then wanted to see whether he could learn about these subjects in advance, during his vacation, he asked his parents to obtain the textbooks for him. Maja says that he forgot about play and playmates; he set to work on the theorems, not by taking the proofs from books, but rather by attempting to prove them for himself. He sat all alone for days, immersed in the search for solutions and proofs, and he often found proofs that were different from those found in the books. Maja reports that during this vacation of a few months from school, Albert independently worked his way through the entire prospective Gymnasium syllabus."[41]

**4. Teaches himself Natural Science and Philosophy**

**4.1 Max Talmud recommends Bernstein's Natural Sciences books**

Maja recalls, that "At the same time the philosophical spirit began to stir in" Albert. Then "A poor Jewish medical student of Polish nationality", Max Talmud, "for whom the Jewish community had obtained free meals with the Einstein family, provided the impetus and thus repaid richly with intellectual stimulation what he received in material benefit. It was he who initiated the youth into the world of philosophical thought."[42] Reiser adds, "Each Thursday his parents invited a poor Russian-Jewish student to dinner. This practice was a form of customary beneficence silently exercised in Jewish circles."[43]

In 1932, Talmud recollected memories from these visits to the Einstein's family home: "Early in the winter of 1889-1890, shortly after I have matriculated as medical student at the University of Munich, I was introduced into the comfortable, cheerful Einstein home. Albert, a pretty dark-haired, brown-eyed boy, was then in the third grade of the Luitpold Gymnasium. Although I was his senior by eleven years, close fellowship soon developed between us, for Albert was able to converse with a college graduate on subjects above the comprehension of children of his age. He showed a particular inclination towards physics and took pleasure in talking about physical phenomena. I gave him therefore as reading matter, A. [Aaron David] Bernstein's *Popular Books on Natural Science* [*Naturwissenschaftlichen Volksbücher*] and L. Büchner's *Force and*

---

[40] Reiser, 1030, p. 37.
[41] Winteler-Einstein, 1924, *CPAE*, Vol. 1, p. xx, p. 1xi.
[42] Winteler-Einstein, 1924, *CPAE*, Vol. 1, pp. xx-xxi, p. 1xii.
[43] Reiser, 1930, p. 36.

*matter* [*Kraft und Stoff*].[44] Each of these books made a profound impression upon the boy, but Bernstein's books, which describe physical phenomena most attractively, had an especially marked influence on Albert and greatly increased his interest in physics."[45]

Reiser described Bernstein's books, "these very popular little books, twenty-one of which were published, had at that time an extraordinary large circulation. They were a gay-colored, beautiful atlas of nature within the limits of child's comprehension".[46] Bernstein's volumes were not involved so much with *theoretical physics*. They were indeed an atlas that presented the wonders of science; and as such the emphasis was put on technical innovations, application of science; new horizons in planetary and comet discoveries, earth science, and Darwin's theory. Bernstein's books supplied scientific answers to children's and lay people's quandaries, and Bernstein often presented the explanations using imaginary fantastic stories.

The first essay of volume 1 is "1. The Goal of Natural Science". Bernstein presents in this chapter different physical phenomena: waves,[47] sound waves (which he explains to be different from light), and light. Bernstein says that the speed of light in space is previously taken as either myth or exaggeration. Bernstein explains about light and the ether: it fills the whole of space, and its wave motions are perceived as light.[48]

Bernstein says that no man has ever seen this ether and no man will ever see it.[49] Bernstein then goes on to explain that the ether is lighter than we would expect and waves propagate quickly in the ether; the velocity of the light surpasses that of sound, and it is almost million times faster. He writes that, the speed of light is 300,000 km per second.[50] Bernstein raised the possibility that all natural forces might propagate with the speed of light in space as well. Everyone was well convinced of the fact, that there are forces in nature which traverse space with almost inconceivable velocity.[51]

---

[44] "I also remember", said Einstein to Seelig, "that at the age of 13 I read with enthusiasm Ludwig Büchner "Force and matter", a book which I later found to be rather childish in its ingenuous realism". Seelig, 1956, p. 12; Seelig, 1954, p. 14.
Reiser wrote, "He was strongly impressed by Büchner's 'Force and Matter', a popular book of the time, but he did not yet perceive its philosophical weakness". Reiser, 1930, p. 38.
[45] Talmay, Max, "Personal Recollections of Einstein's Boyhood and Youth", *Scripta Mathematica*, New York, 1932, Vol. 1, pp. 68-71; pp. 68-69.
[46] Reiser, 1930, p. 36; Bernstein, Aaron, *Naturwissenschaftliche Volksbücher: Wohlfeile Gesammt-Ausgabe*, 1870/1897, Berlin: Ferd. Dümmlers Berlagsbuchhandlung.
[47] Bernstein, 1870/1897, Erster Teil, pp. 3-7.
[48] "Wenn man sonst von der Geschwindigkeit sprach, mit welcher das Licht die Räume durchfliegt, so hielten es viele für eine Fabel oder eine wissenschaftliche Übertreibung. Jetzt, wo man täglich Gelegenheit hat, die Geschwindigkeit des elektrischen Stromes am elektromagnetischen Telegraphen zu bewundern, jetzt leuchtet es auch wohl allen ein, dass es Naturkrafte gibt, die in unbegreiflichen Geschwindigkeiten sich durch den Raum fortpflanzen". Bernstein, 1870/1897, Erster Teil, pp. 7-10.
[49] "Man hat ihm einen Namen gegeben, trotzdem ihn niemals ein Mensch gesehen hat oder je sehen wird: er heisst der Äther".
[50] Bernstein, 1870/1897, Erster Teil, p. 11.
[51] Bernstein, 1870/1897, Zweiter Teil, chapters LV-LVII.

Starting in the Third volume [Dritter Teil] Bernstein began to explain electricity and magnetism.[52] In Bernstein's volume 8 there is an essay "On the Rotation of the Earth", the first chapter of which is entitled "Die Uhr" (The clock). Bernstein starts with an explanation of our tacit notion of a clock from daily life, and then relates it to the daily rotation of the earth.[53] He then describes the pendulum,[54] and describes the machinery of the pocket watches that causes the "Tick-Tock"; finally he discusses in a few pages the rotation of the earth.[55] Then Bernstein arrives at the essay, "On The Speed of Light", the first chapter of which is On Light". Bernstein openes the essay by declaring that, "light travels four thousand miles per second!".[56] He then discusses the measurement of the velocity of light by Jupiter satellites,[57] and arrives at Bradley's attempts at measuring the velocity of light.[58]

In the chapter discussing Bradley's measurements of the velocity of light, Bernstein imagined a thought experiment:[59]

Imagine a wagon, the walls of which are punched at both sides by a bullet. The bullet is shot through the wagon. The man who shot the bullet directed his gun so that the bullet would be shot directly across the wagon. However, investigation of the holes punched in the wagon shows that the bullet was shot not exactly across the wagon in a straight line, but a little oblique. The holes punched through the walls of the wagon, as the ball enters the wagon, show that the bullet enters a bit forward. However, anyone who had seen this shot from inside the wagon would have claimed that the shot must have been impossible: he would claim that, the bullet was deliberately shot obliquely, and not to the front. And yet the shot was directed straightly, and the bullet was shot to the front, although it was seen to go through the wagon in an oblique direction. Bernstein then gives the explanation for this: the wagon was moving while the bullet was piercing the first wall, and it still continued to move while the bullet passed through the wagon until it pierced the other wall. During the time that the ball was passing from one wall to the other, the wagon moved a little ahead. And thus the shot at the opposite wall could not be the same as it would have been were the wagon was at rest. This is aberration.

The German historian of science, Friedrich Herneck analyzed Bernstein's first volume. In the chapter "First encounter with the speed of light" of Herneck's biography of Einstein, he quoted from Bernstein's first volume.[60] According to Herneck, Einstein's later thoughts about the speed of light could be influenced by his earlier reading of Bernstein's popular science books. Herneck suggested that Einstein might have thought of the speed of light already in Munich when he was twelve years old.[61] Einstein did not pronounce anything on this matter.

---

[52] Bernstein, 1870/1897, Dritter Teil.
[53] Bernstein, 1870/1897, Achter Teil, pp. 100-104.
[54] Bernstein, 1870/1897, Achter Teil, pp. 104-110.
[55] Bernstein, 1870/1897, Achter Teil, pp. 110-127.
[56] "Das licht bewegt sich vierzigtausend Meilen in einer Sekunde!". Bernstein, 1870/1897, Achter Teil, p. 128.
[57] Bernstein, 1870/1897, Achter Teil, pp. 133-140.
[58] Bernstein, 1870/1897, Achter Teil, pp. 140-146.
[59] Bernstein, 1870/1897, Achter Teil, pp. 141-142.
[60] Herneck, Friedrich, *Albert Einstein: ein Leben für Wahrheit, Menschlichkeit und Frieden*, 1963, Berlin: BuchVerlag der Morgen, p. 32.

In volume sixteen Bernstein describes the wonders of the skies, and then dedicates a chapter to each planet; finally he invites his readers to join him for a fantasy journey into space. Under the title, "Eine Phantasie-Keise im Weltall", "1. Die Abreife", Bernstein described his imaginary journey: Suppose you want to perform a voyage to space. You need a passing-card, and some provisions, food, a suitcase. Although our voyage is going to be very fast, we are going deep into space. In our suitcase we will take our thoughts. "We travel by water? On the back of the horse? By train? None of that! We travel with the help of an electrical telegraphic apparatus!" [62]

This could have influenced Einstein on his way to special relativity. Herneck thought that Bernstein might have inspired Einstein when he propounded his Aarau thought experiment of him chasing a light beam.[63] Indeed a few years later, Einstein at school, in Aarau, imagined a journey on a light beam as well (not exactly on a telegraphic signal); a thought experiment of him chasing a light beam,[64]

"[…] a paradox upon which I had already hit at the age of sixteen: If I pursue a beam of light with a velocity $c$ (velocity of light in a vacuum), I should observe such a beam of light as an electromagnetic field at rest through spatially oscillating."

At the same time Einstein was immersed with music. In 1940, Einstein told Frank,

"I really began to learn [violin] only when I was about 13 years old, mainly after I had fallen in love with Mozart's sonatas. The attempt to reproduce, to some extents, their artistic content and their singular grace compelled me to improve my technique, which improvement I obtained from these sonatas without practicing systematically. I believe, on the whole, that love is a better teacher than sense of duty – with me, at least, it certainly was".[65]

**4.2 Einstein reads a small geometry book**

In addition to Bernstein's books, Einstein also read a geometry book and taught himself geometry. Max Talmey says that "contrary to popular belief he [Albert] had an unusual predilection for mathematics, and because of this fact I gave him, after his promotion to the fourth grade, Spieker's textbook on geometry".[66]

---

[61] Herneck, 1963, p. 50. Bernstein, 1870/1897, Achter Teil, p. 148.
[62] Bernstein, 1870/1897, Sechzehnter Teil, p. 54.
[63] Herneck, p. 50.
[64] Einstein, 1949, pp. 48-51.
[65] Hoffmann and Dukas, 1973 p. 20; *CPAE*, Vol. 1, editorial note. 39, p. 1viii.
[66]Talmud gave Einstein Theodor Speiker's *Lehrbuch der ebenen geometrie mit übungsaufgaben für höhere lehranstalten*, a popular textbook from 1890. Talmey, 1932, p. 69.

Reiser recounts, "A strong, indelible, and creative impression not to be overlooked was produced on the twelve-year-old boy by the first small geometry book held in his tender hands. Geometry was not yet an assigned study in the secondary school. That did not begin until shortly after. But Albert already possessed the textbook, which caused him a tremendous excitement." [67] Was "the first small geometry book" Spieker's geometry book?

Recall that after the wonder that Einstein had experienced at the age of five with the compass, the geometry book was considered as a second wonder: "At the age of 12 I experienced a second wonder of a totally different nature: in a little book dealing with Euclidean plane geometry, which came into my hands at the beginning of a schoolyear".[68] However, we still do not know which geometry book was considered as wonder.

Talmey recounts, "I used to visit his home every week and whenever I came he delighted in showing me his solutions of new problems which he had found in the book [Spieker's geometry book …] and soon he had mastered the whole textbook. He then turned to higher mathematics [...]. His progress in mathematics was so rapid that very soon I was no longer a match for him in the subject."[69]

According to Maja, Albert's Uncle Jacob, an engineer who had a comprehensive mathematical education, posed difficult mathematical problems for Albert. Albert invariably found a correct solution, and even found an entirely original proof for the Pythagorean theorem. When he obtained such results, Albert "was overcome with great happiness", says Maja, "and was already then aware of the direction in which his talents were leading him".[70]

Moszkowski reported,[71]

"The boy plunged himself for three weeks into the task of solving the Pythagorean theorem, using all his power of thought. He came to consider similarity of triangles (by dropping a perpendicular from one vertex of the right-angled triangle on to the hypotenuse), and was thus led to a proof for which he had so ardently longed!"

This style of work "considering similarity" of two objects or things was going to be a heuristic guide in Einstein's later work.

Reiser repeated Moszkowski and Maja's above descriptions, "This uncle also told him of the Pythagorean theorem […] Soon the small textbook was his favorite reading. It brought him classical experience of perfect harmony, and his preoccupation with mathematics became the most beautiful adventure of his youth. When he secured

---

[67] Reiser, 1930, p. 35.
[68] Einstein, 1949, p. 10.
[69] Talmey, 1932, p. 69.
[70] Winteler-Einstein, 1924, *CPAE*, Vol. 1, p. xx, pp. 1xi-1xii.
[71] Moszkowski, 1921a, pp. 224-225; Moszkowski, 1921b, pp. 222-223.

Spieker's geometry, he at once succeeded in solving all the exercises, including the most difficult, with the exception of two or three". Reiser explained that Einstein worked through Speiker's geometry book, but only after he read "the small geometry book".[72]

The small geometry book, which was considered as wonder was probably the textbook that Albert received from his uncle Jacob, Adolf Sickenberger, *Leitfaden der Elementaren Mathematik, Zweiter Teil, Planimetrie*, 1888. This book was issued in three separate parts, and part 2 fits the description Einstein later gave of "Holly Geometry Book"[73] that he received at the age of 12 at the beginning of the school year.[74]

Einstein recounted in his *Autobiographical Notes*,[75]

"At the age of twelve through sixteen I familiarized myself with the elements of mathematics, including the principles of differential and integral calculus. In doing so I had the good fortune of encountering books that were not too particular regarding logical rigor, but that permitted the principal ideas to stand out clearly. This occupation was, on the whole, truly fascinating."

**4. 3 Max Talmud recommends Kant**

After science and mathematics Talmud went on to philosophy. He recommended to Einstein Kant's *Critiques of Pure Reason*: "Philosophy then became a frequent subject of our conversations. I recommended to him the reading of Kant, although he was then only thirteen years old; but even at that age, the works of this master, difficult as they are to most readers, seemed to be clear to him and thereafter Kant became his favorite philosopher."[76]

Einstein had read all three main works of Kant before he was 16: *Kritik der reinen Vernunft* (*Critique of Pure Reason*), *Kritik der praktischen Vernunft* (*Critique of Practical Reason*), and *Kritik der Urteilskraft* (*Critique of Judgment*). Einstein said,

---

[72] Reiser, 1930, pp. 35-36.
[73] "das heilige Geometrie-Büchlein".
[74] *CPAE*, Vol. 1, note 49, p. lxi.
[75] Einstein, 1949, p. 13;
Lewis Pyenson writes in 1985 about mathematical books that Einstein gave in his will to his secretary Helen Dukas, "According to correspondence from Miss Dukas, the four books are, respectively: Georg Freiherr von Vega, *Logarithmisch-Trigonometrisches* Handbuch ed C Bremiker (Berlin 1869); H B Liibsen, *Ausführliches Lehrbuch der Analysis zum Selbstunterricht mit Rücksicht auf die Zwecke des praktischen Lebens* (Leipzig 1868); H B Liibsen, *Einleitung in die Infinitesimal-Rechnung (Differential und Integral-Rechnung) zum Selbstunterricht . . .* (Leipzig 1869); Heis and Eschweiler, *Lehrbuch der Geometrie zum Gebrauch an höheren Lehranstalten* 2: *Stereometrie* (Cologne 1881). Miss Dukas declined to let me examine the books, which until her recent death were located at the Institute for Advanced Study, Princeton". Pyenson, Lewis, *The Young Einstein: The Advent of Relativity*, 1985, Boston: Adam Hilger, pp. 29-30.
[76] Talmey, 1932, p. 69.

"'I did not grow up in the tradition of Kant', he maintained, 'and only later understood the valuable material which is hidden in his teaching' ".[77]

## 5 Einstein is Free in Italy

The Einstein's firm in Munich was liquidated and was moved to Italy. Reiser says that Einstein's "father suffered one mishap after another in his business undertakings. The electrical engineering plant did not prosper. He found it impossible to make his living."[78] When a great promise was shown in Italy, the Italian representative of the firm proposed moving the plant to Italy. Albert's ambitious uncle, Jakob Einstein, was at once taken with the idea and he persuaded Albert's father, Hermann Einstein, to make the change, literally sweeping him along.[79] In the summer of 1894, the plant was transferred to Pavia. Albert's parents, his sister and uncle Jakob all moved first to Milan in 1894 and a year later to Pavia.

Maja recalls with much agony, "The firm in Munich was liquidated. The lovely estate with the villa in which Albert Einstein had spent a happy childhood was sold to a building constructor, who immediately turned the handsome grounds into a construction site, cutting down the magnificent old trees and erecting an entire row of ugly apartment houses. Until the time of their move the children had to watch from the house as these witnesses to their most cherished memories were destroyed".[80]

Reiser writes, "They [Albert's parents] moved to Milan, while Albert remained alone at Munich to complete his schooling. Registered at the pension of a friendly old lady, he led an extremely quiet existence".[81] Maja also says, that when the family moved to Italy in 1894, they decided to leave young Albert in Munich to ensure he would graduate school and finish the last three years of the Gymnasium; he boarded with a family in Munich, while relatives took care of him. Another reason was the Italian language, which was foreign to him.[82]

---

[77] Carl Seelig writes, "Although in his Gymnasium years he studied Kant, the German metaphysicians did not particularly influence him". Seelig, 1956, p. 114; Seelig, 1954, p. 135. Later he read Kant again when he studied at the Zürich Polytechnic. There he enrolled in Prof. Dr. August Stadler's course, which was delivered during the summer semester of 1897 "The theory of scientific thought – Kantian philosophy". Seelig, 1956, p. 25; Seelig, 1954, p. 30. While a professor at Prague, in his leisure time Einstein often appeared with his violin in the salon of Frau Bertha Fanta, a most cultivated woman. In her salon chamber music was played under the leadership of the Zionist philosopher Hugo Bergmann and until long after midnight Kant's *Critique of Pure Reason* and other philosophical writings were read. Seelig, 1956, p. 121; Seelig, 1954, p. 146.
[78] Reiser, 1930, p. 38. Herman's younger brother Jakob finished his studies in engineering and started the plumbing and electrical business in Munich. Since he wanted money from Hermann's wife's family, he prevailed upon his brother Hermann to join in the venture, both personally as business manger and with large investment. They both inaugurated the J. Einstein & Co. Electrotechnical Factory in Munich. Winteler-Einstein, 1924, *CPAE*, Vol. 1, p. xvi, p. 1i.
[79] Winteler-Einstein, 1924, *CPAE*, Vol. 1, pp. xvi-xvii, pp. 1ii-1iii.
[80] Winteler-Einstein, 1924, *CPAE*, Vol. 1, p. xvii, p. 1iii.
[81] Reiser, 1930, p. 40.
[82] Winteler-Einstein, 1924, *CPAE*, Vol. 1, p. xxi, pp. 1xiii.

Reiser said that when Einstein was a very young boy, "Sophisticated small cousins came from Genoa, told of Italy and its people, of harbors, ships and sailors. With old factory chests they played 'going to the sea'. The boy [Albert's] world grew larger."[83] Albert was now fifteen, alone and, says Reiser, "Italy, however, seemed a paradise to him. His younger cousins from Genoa had already told him much about it. Now his parents lived there and their letters painted for him this land of sun, color, and free, natural people."[84] And, "he felt miserable at having to occupy himself with things in which he was not interested but which he was supposed to learn only because he had to take an examination in them. This feeling of dissatisfaction grew greater when his parents departed and left him in a boarding-house. He felt himself a stranger among his fellow students […]."[85]

Here is how Einstein described the situation in a letter written in 1940:[86]

"When I was in the seventh grade at the Luitpold Gymnasium [and thus about fifteen] I was summoned by my home-room teacher[87] who expressed the wish that I leave the school. To my remark that I had done nothing amiss he replied only 'your mere presence spoils the respect of the class for me'.

I myself, to be sure, wanted to leave school and follow my parents to Italy. But the main reason for me was the dull, mechanized method of teaching. Because of my poor memory for words, this presented me with great difficulties that it seemed senseless for me to overcome. I preferred, therefore, to endure all sorts of punishments rather than learn to gobble by rote."

The rebel and miserable Einstein who was intolerable in school, and would learn only topics he liked, wanted to leave. His teacher who demanded obedience could not withstand his presence in class.

Indeed Albert's situation at school was miserable, as Maja says. In addition he longed for home and his parents. At school he felt that the style of teaching was repugnant, militaristic; he could not withstand "the systematic training in the worship of authority that was supposed to accustom pupils at an early age to military discipline". He could not even think about the moment he would have to wear a soldier's uniform in order to fulfill his military duty.[88]

According to the then German citizenship laws, a male citizen must have emigrated by the age 16; otherwise, if he failed to report the military service, he would have

---

[83] Reiser, 1930, p. 28.
[84] Reiser, 1930, p. 40.
[85] Frank, 1947/2002, p. 16; Frank, 1949/1979, p. 30.
[86] Hoffmann and Dukas, 1973, p. 25.
[87] The teacher Joseph Degenhart who had prophesied that Einstein would never amount to anything. *CPAE*, Vol. 1, notes 56 and 58, p. lxiii.
[88] Winteler-Einstein, 1924, *CPAE*, Vol. 1, p. xxi, pp. 1xiii.

been declared a deserter. For this reason Albert decided to leave Germany as quickly as possible.[89]

Einstein was depressed and nervous; he thus searched for a way out of school. Hence when the professor in charge of his class – the one who expressed the wish that Einstein would leave the school – annoyed him again, Albert decided he would leave school. He obtained a certificate from the family doctor, presented it to the school principal, withdrew from the school on December 29 1895, and off he went by train which crossed the Alps straight to Italy, Milan to his parents.[90]

Arriving at his parents' home in Milan, Einstein announced he would not return to Munich, nonetheless he promised his stunned parents that he was going to independently prepare himself for the entrance examination to the Zürich Polytechnic School (*Eidgenössische Polytechnische Schule*) in autumn.[91] Einstein used to call this school the *Züricher Polytechnikum*, the "Poly" in Zürich. This institute was one of the best teaching and research institutes in all Europe. Einstein purchased the three volumes of the advanced large textbook by Jules Violle, *Lehrbuch der Physik*, and worked through nearly all of them and prepared for exams.[92]

Maja says, "his work habits were rather odd: even in a large, quite noisy group, he could withdraw to the sofa, take pen and paper, in hand, set the inkstand precariously on the armrest, and lose himself so completely in a problem that the conversation of many voices stimulated rather than disturbed him; an indication of remarkable power of concentration".[93]

On arriving at Milan, Einstein "told his father that he wanted to renounce his German citizenship. His father, however, kept his own, so that the situation was rather unusual. Also since Einstein could not acquire any other citizenship immediately, he became stateless. Simultaneously he renounced his legal adherence to the Jewish religious community".[94] Einstein had thus become a person without a religion affiliation (*Konfessionslos*). When young Albert gave up his German citizenship, he was no longer subject to the German army and he was stateless until he was much later a Swiss citizen.

*I wish to thank Prof. John Stachel from the Center for Einstein Studies in Boston University for sitting with me for many hours discussing special relativity and its history. Almost every day, John came with notes on my draft manuscript, directed me to books in his Einstein collection, and gave me copies of his papers on Einstein, which I read with great interest.*

---

[89] Winteler-Einstein, 1924, *CPAE*, Vol. 1, p. xxii, pp. lxiv.
[90] Winteler-Einstein, 1924, *CPAE*, Vol. 1, p. xxi; *CPAE*, Vol. 1, note 58, pp. lxiii.
[91] Winteler-Einstein, 1924, *CPAE*, Vol. 1, p. xxi, pp. lxiv.
[92] Winteler-Einstein, 1924, *CPAE*, Vol. 1, p. xxii, p. lxiv.
*Erster Theil. Mechanik, Erster Band. Allgemeine Mechanik und Mechanik der festen Körper* (general mechanics and the mechanics of the rigid body)/ *Zweiter Theil. Akustik und Optik,* and also the second volume: *Geometrische Optik* (acoustics and optics, geometric optics)/ *Zweiter Theil. Mechanik der flüssigen und gasförmigen* (mechanics of liquid and gases).
[93] Winteler-Einstein, 1924, *CPAE*, Vol. 1, p. xxii, p. lxiv.
[94] Frank, 1947/2002, p. 17; Frank, 1949/1979, p. 34.